\documentclass{aa}  

\usepackage{graphicx}
\usepackage{txfonts}
\usepackage{hyperref}
\usepackage{siunitx}
\usepackage{color}

\usepackage{amsmath}

\def\zsource{z_{\mathrm{s}}}
\def\zlens{z_{\mathrm{g}}}
\def\sflux{f_{\mathrm{s}}}

\def\kappaext{\kappa_{\mathrm{ext}}}

\def\tein{\theta_{\mathrm{Ein}}}
\def\tcirc{\theta_{\mathrm{circ}}}

\def\Aquad{A_{\mathrm{quad}}}
\def\Aquadobs{A_{\mathrm{quad}}^{(\mathrm{obs})}}
\def\AquadSIE{A_{\mathrm{quad}}^{(\mathrm{SIE})}}

\def\psilens{\boldsymbol{\psi}_\mathrm{g}}

\def\prlens{{\rm P}_\mathrm{g}}
\def\prquad{{\rm P}_\mathrm{quad}}
\def\prsource{{\rm P}_\mathrm{s}}

\def\pdet{{\rm P}_\mathrm{det}}

\def\crosssect{\sigma_\mathrm{{SL}}}
\def\crosssecteff{\sigma_\mathrm{{SL}}^{(\mathrm{eff})}}

\def\lmst{\lambda_{\mathrm{mst}}}

\def\Sref#1{Section~\ref{#1}\xspace}
\def\Fref#1{Figure~\ref{#1}\xspace}

\def\Eref#1{Equation~\ref{#1}\xspace}

\def\pr{{\rm P}}

\defcitealias{Son++23}{S23}
\setcitestyle{notesep={}}

\begin{document}

   \title{Caustic area biases and how to deal with them}
   \titlerunning{Caustic area biases}
   \authorrunning{Sonnenfeld}

   \author{Alessandro Sonnenfeld\inst{\ref{sjtu1},\ref{sjtu2},\ref{sjtu3}}
          }

   \institute{
Department of Astronomy, School of Physics and Astronomy, Shanghai Jiao Tong University, Shanghai 200240, China\\
              \email{sonnenfeld@sjtu.edu.cn}\label{sjtu1} \and
Shanghai Key Laboratory for Particle Physics and Cosmology, Shanghai Jiao Tong University, Shanghai 200240, China\label{sjtu2} \and
Key Laboratory for Particle Physics, Astrophysics and Cosmology, Ministry of Education, Shanghai Jiao Tong University, Shanghai 200240, China\label{sjtu3}
             }

   \date{}

 
  \abstract
    {
Quadruply-imaged strongly lensed quasars (quads) are routinely used for measurements of the expansion rate of the Universe with time delays.
It has recently been suggested that any quad lens is subject to a Malmquist-like bias that causes the inferred area enclosed within the tangential caustic to be systematically underestimated, and that such a bias might translate into a corresponding bias on the expansion parameter.
In this work we extended that analysis by also considering the effect of Eddington bias. 
We found that the sign and amplitude of the combined bias depend on the functional form of the caustic area distribution of the lens population and on the noise associated with the caustic area estimation process.
Based on simulations, we estimated that the corresponding impact on $H_0$ is on the order of a percent or smaller.
If the likelihood of the lensing data is known, then the bias can be accounted for when modelling the lens population.
However, ignoring the criteria used to select a quad might lead to a bias at the lens modelling stage that causes the inferred caustic area to be overestimated.
Such a bias disappears for lens models that are well constrained by the data.
}
   \keywords{
             Gravitational lensing: strong 
               }

   \maketitle
%

\section{Introduction}\label{sect:intro}

Strong gravitational lensing is a formidable tool for the study of galaxy structure and cosmology.
Time delay measurements on strongly lensed variable sources, such as quasars or supernovae, can be used to constrain the Hubble-Lema\^{i}tre constant $H_0$ independently of local distance calibrations \citep{Ref64,Suy++17,Mil++20,Bir++20,TSM22}.
As the precision of time-delay analyses improves \citep{Sha++23,Ace++23,Kel++23}, it becomes increasingly important to keep systematic uncertainties under control. 
The accuracy of a time-delay lensing inference depends critically on how well the mass distribution of the light-deflecting system is characterised, including both the main lens and mass along the line of sight \citep{S+S13,Xu++16,Son18,Koc20}.
While great effort has been put into developing methods to constrain both these components \citep{Gre++13,Sha++18,Yil++23,Wel++23,Par++23}, it is important to make sure that no potential source of error is left unchecked \citep[][]{Koc21,VdV++22,Gom++22,Gom++23}.

Recently, \citet{B+S24} pointed out the existence of a selection effect that might bias inferences of $H_0$ from time-delay lensing.
\citet{B+S24} focused on quadruply-imaged point sources (quads), which make up the majority of systems used so far for time-delay analyses.
In order for a source to be lensed into a quad, it is necessary for it to lie in the region of the source plane enclosed within the tangential caustic of the lens.
\citet{B+S24} argued that, since galaxies with an intrinsically larger tangential caustic area are more likely to be strong lenses, the selection of quads introduces a bias, as a result of which the estimated caustic area is systematically smaller than the true one.
This bias was said to be similar in nature to the Malmquist bias, in which the observed brightnesses of a flux-limited sample of stars or galaxies are systematically overestimated near the detection limit.
If present, such a bias would translate into a corresponding bias on $H_0$.

Here we build upon the work of \citet{B+S24} to understand more generally how uncertainties on the estimated caustic area impact the inference of other lens properties.
First, we derive an expression for the caustic area bias of a lens survey, and apply it to a few example cases. 
We show that, due to an Eddington bias effect, the sign of the bias at the high mass end is opposite than that due to the Malmquist bias alone.
Second, we focus on the lens modelling stage and show how a seemingly innocuous assumption can introduce a positive bias on the caustic area.
Along with the examples of biases, we suggest strategies for removing them.

This paper is organised as follows. In \Sref{sect:lensing} we briefly introduce basic notions on strong lensing. 
In \Sref{sect:population} we introduce the problem of the caustic area bias and derive an expression for its quantification.
In \Sref{sect:modelling} we show an instance of a caustic area bias introduced when modelling an individual lens and explain how to eliminate it.
In \Sref{sect:concl} we discuss our results and draw conclusions.


\section{Strong lensing basics}\label{sect:lensing}

A strong lens can be fully described by specifying its dimensionless surface mass density $\kappa$ as a function of position in the sky $\boldsymbol\theta$,
\begin{equation}
\kappa(\boldsymbol\theta) \equiv \frac{\Sigma(\theta)}{\Sigma_{\mathrm{cr}}},
\end{equation}
where $\Sigma$ is the physical projected mass density and $\Sigma_{\mathrm{cr}}$ is the critical surface mass density for lensing, determined by the redshift of the lens and of the background source.
Given $\kappa$, we can compute the deflection angle as a function of position:
\begin{equation}
\boldsymbol\alpha(\boldsymbol\theta) = \frac{1}{\pi}\int_{\mathbb{R}^2} d^2\boldsymbol\theta' \kappa(\boldsymbol\theta')\dfrac{\boldsymbol\theta - \boldsymbol\theta'}{|\boldsymbol\theta - \boldsymbol\theta'|^2}.
\end{equation}
Images of a source at angular position $\boldsymbol\beta$ form at the solutions of the lens equation
\begin{equation}
\boldsymbol\beta = \boldsymbol\theta - \boldsymbol\alpha(\boldsymbol\theta).
\end{equation}
The magnification of lensed images can be obtained by means of the Jacobian matrix of the lens equation, $A\equiv \partial\boldsymbol\beta/\partial\boldsymbol\theta$. Lenses can have up to two curves where the $A$ matrix is singular, one for each eigenvalue. These are called critical curves and correspond to points of formally infinite magnification.
In our work we will consider exclusively lenses with elliptically symmetric mass distributions. For such lenses, the two critical curves are labelled tangential and radial, depending on whether the corresponding magnification is largest in the direction tangential or perpendicular to the isodensity curves.
The curve obtained by mapping a critical curve to the source plane with the lens equation is called caustic.

Throughout this work we focus on a particular class of lenses, with elliptically symmetric projected mass distribution and a power-law radial density profile. Their dimensionless surface mass density can be described by
\begin{equation}\label{eq:plkappa}
\kappa(\tcirc) = \frac{3-\gamma}{2}\left(\frac{\tcirc}{\tein}\right)^{1-\gamma},
\end{equation}
where $\tcirc$ is the circularised radius,
\begin{equation}
\tcirc \equiv \sqrt{q\theta_1^2 + \frac{\theta_2^2}{q}},
\end{equation}
$q$ is the axis ratio, $\tein$ is the Einstein radius and $\gamma$ the corresponding 3-dimensional density slope.

\begin{figure*}
\includegraphics[width=\textwidth]{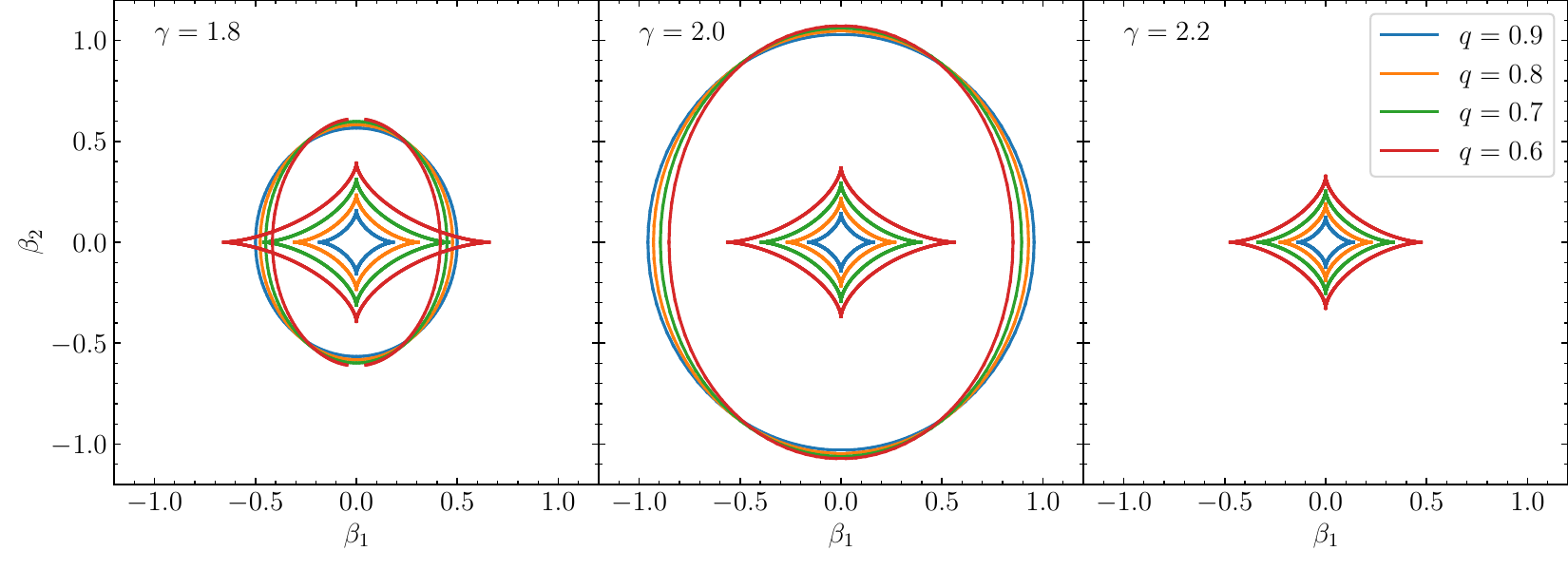}
\caption{
Caustics of power-law lenses with elliptical mass distributions.
Each panel corresponds to a value of the density slope $\gamma$. Different colours correspond to different values of the axis ratio, as indicated in the legend. The outer, elliptical-shaped curve is the radial caustic (in the $\gamma=1.8$ case) or the cut (in the $\gamma=2$ case), while the inner diamond-shaped one is the tangential caustic.
The major axis of the lens points in the horizontal direction.
Caustics were obtained with the software {\sc Glafic} \citep{Ogu21}.
\label{fig:caustics}
}
\end{figure*}

\Fref{fig:caustics} shows caustics of elliptical power-law lenses, with different values of $q$ and $\gamma$.
Starting from the $\gamma=1.8$ case (left-hand panel), the outer, elliptical-shaped, curve is the radial caustic, while the diamond-shaped one is the tangential caustic.
The position of the source relative to the caustics determines the number of images that are produced. If the source is inside both caustics, then five images are formed. One of these images is, in all practical cases, extremely de-magnified and hence ignored: the end result is a quad lens.
At fixed lens properties, the source-plane area that is mapped to a quad configuration generally consists of the region enclosed within the tangential caustic. The only exception is when the tangential and radial caustic intercept each other, such as in the $q=0.6$, $\gamma=1.8$ example. 
In those cases, only the region enclosed by both caustics gives rise to sets of four images.

Power-law ellipsoid lenses with $\gamma \geq 2$ have no radial critical curve, and therefore no radial caustic.
In the limit $\gamma \rightarrow 2$, the radial caustic turns into a cut, which is a curve that separates regions of the source plane that are mapped into one or two images.
Differently from a caustic, sources that lie exactly on the cut do not give rise to infinitely magnified images.
The cut is visible in \Fref{fig:caustics} for the $\gamma=2.0$ case (middle panel), while is outside of the plotted range in the $\gamma=2.2$ case (right-hand panel). Sources that lie in the region between the tangential caustic and the cut are mapped into two images, while those inside the tangential caustic are mapped into four images.

\Fref{fig:caustics} shows a clear trend between the axis ratio of the lens mass distribution and the area enclosed by the tangential caustic: the higher the ellipticity, the larger the area. 
Although more difficult to see, the tangential caustic area also decreases with increasing $\gamma$, at fixed $q$. 
A larger or smaller tangential caustic area generally translates into a higher or lower probability of a lens creating a quad configuration.
However, the cross-section for a quad event depends also on the intrinsic brightness of the source and the magnification of the multiple images, because the sole fact that four images are produced does not guarantee that they are detectable.
For the sake of simplicity, we ignore magnification throughout this work. This assumption corresponds to the limit of a very bright source, which is a reasonable one in the case of lensed quasars.
In \Sref{sect:concl} we discuss how the findings of this paper can be generalised to the case of fainter sources.

The mass distribution of a lens can be inferred by fitting a model to the observed lensing data, such as the positions of the multiple images. This is a fundamental step in all time-delay lensing analyses.
However, there is a limit to how much information can be extracted in the process. 
Given a lens with surface mass density distribution $\kappa(\boldsymbol\theta)$, a transformation of the kind
\begin{align}
\kappa(\boldsymbol\theta) & \rightarrow \lmst\kappa(\boldsymbol\theta) + 1 - \lmst,\label{eq:kappatrans} \\
\boldsymbol\beta & \rightarrow \lmst\boldsymbol\beta\label{eq:sourcetrans},
\end{align}
leaves image positions and flux ratios unchanged, while varying the time delay between images.
The original and the transformed lens are therefore indistinguishable on the basis of the data typically used to constrain lens models, but different values of $H_0$ are needed to match a given time delay. 

\Eref{eq:sourcetrans} correspond to a uniform rescaling of the source plane coordinates. This means that the area enclosed within any caustic is a factor $\lmst^2$ of the original one. In the bright source limit, the probability of creating a quad event is then also rescaled by the same factor.
The pair of equations \ref{eq:kappatrans} and \ref{eq:sourcetrans} is referred to as a mass-sheet transformation, and the invariance of lensing observables under it is called mass-sheet degeneracy \citep{FGS85}.
As we discuss in the next section, it plays an important role in the caustic area bias.
For a more thorough introduction to strong lensing we refer the reader to \citet{Men21} and references within.


\section{Population-level bias}\label{sect:population}

We consider a population of foreground galaxies, all at the same redshift $\zlens$, and a population of bright background point sources, also at a single redshift $\zsource > \zlens$.
Each lens has an associated caustic area for the formation of quads, $\Aquad$. The distribution in quad caustic area of the population of foreground galaxies is described by a probability distribution $\prlens(\Aquad)$.
Since the probability of a background source being lensed into a quad by a given foreground galaxy is proportional to $\Aquad$, the distribution in $\Aquad$ of the population of quad lenses is
\begin{equation}
\prquad(\Aquad) \propto \Aquad \prlens(\Aquad).
\end{equation}

Next we consider observations. 
Although the caustic area is not a quantity that is usually explicitly fitted for, it is an implicit by-product of any lens model: given a surface mass density distribution inferred from lensing data, it is always possible to calculate the corresponding value of $\Aquad$.
We label the observed $\Aquad$ obtained via lens modelling as $\Aquadobs$.
In general, if the lens model is not a perfect representation of the truth, $\Aquadobs$ will be different from $\Aquad$. Vice-versa, any difference between the inferred and the true value of the quad caustic area is a sign of the model being inaccurate.
With this definition, we introduce the likelihood of measuring a value $\Aquadobs$ given the true value, $\pr(\Aquadobs|\Aquad)$. 
The form of this distribution depends on the details of how $\Aquadobs$ is obtained and on the noise properties of the data.

At fixed observed $\Aquadobs$, the expectation value of the true quad caustic area is simply
\begin{equation}\label{eq:bias}
\rm{E}[\Aquad] = \frac{\int d\Aquad \Aquad\prquad(\Aquad)\pr(\Aquadobs|\Aquad)}{\int d\Aquad \prquad(\Aquad)\pr(\Aquadobs|\Aquad)}.
\end{equation}
Any difference between this quantity and $\Aquadobs$ is a bias on the caustic area.
We label this the population-level bias, to distinguish it from a separate effect studied in \Sref{sect:modelling}.
If the likelihood is symmetric around the true value, then what determines the sign of the bias is the shape of the distribution in quad caustic area of the lenses, $\prquad(\Aquad)$.
In the hypothetical, and unrealistic, case of a uniform distribution in the quad caustic area of the foreground galaxies, $\prlens(\Aquad) \sim \mathrm{constant}$, then $\prquad(\Aquad) \propto \Aquad$ and $\mathrm{E}[\Aquad]$ is larger than the observed value: 
the quad caustic area is underestimated.
This is the scenario implicitly studied by \citet{B+S24}, who did not consider the distribution in caustic area in their treatment.
In this scenario, the root of the bias is uniquely in the presence of an observational scatter and the preferential selection of lenses with larger caustic area, an effect referred to as a Malmquist-type bias.
In general, though, $\prlens(\Aquad)$ is a steeply declining function of $\Aquad$. As a result, the bias can have the opposite sign.

\begin{figure}
\includegraphics[width=\columnwidth]{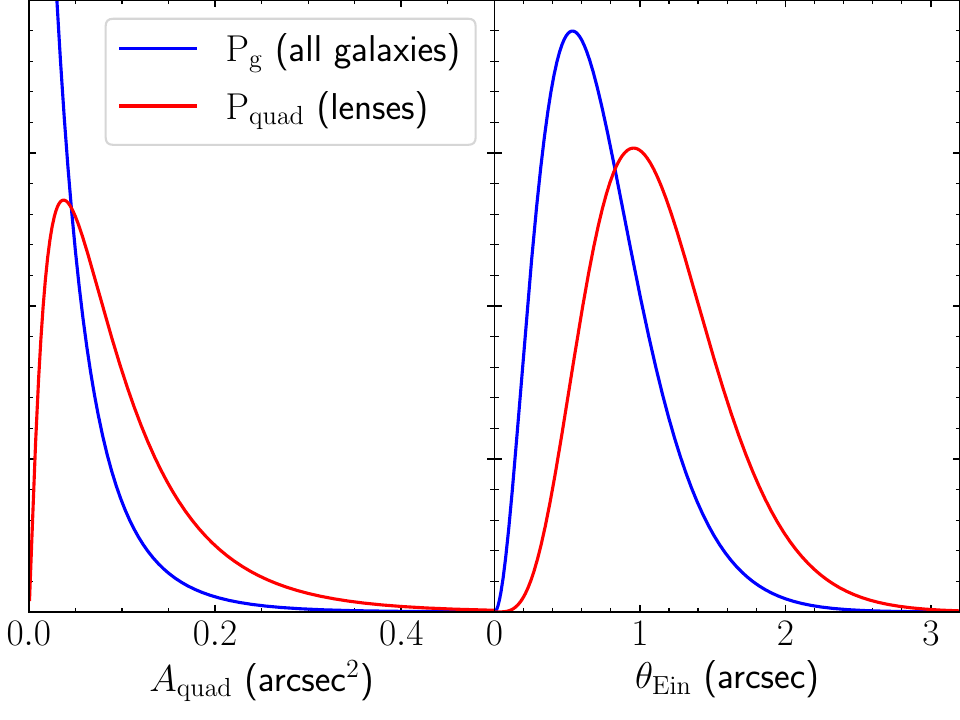}
\caption{
Simulation of foreground galaxies (blue curves) and quad lenses (red curves).
Left: distribution in quad caustic area. Right: distribution in Einstein radius.
All foreground galaxies are singular isothermal ellipsoids with axis ratio $q=0.7$.
\label{fig:dists}
}
\end{figure}

In order to gain intuition on the problem, we carried out the following experiment.
We considered a population of singular isothermal ellipsoid (SIE) lenses. These are elliptical power-law mass distributions with a value of the slope $\gamma=2$. We fixed the value of the axis ratio to $q=0.7$, for simplicity.
With this choice, the quad caustic area is
\begin{equation}\label{eq:aquadsie}
\AquadSIE(q=0.7) = 0.066\left(\frac{\tein}{1''}\right)^2\,\rm{arcsec}^2.
\end{equation}
We then used the following relation, valid for a singular isothermal sphere, to link the Einstein radius to the velocity dispersion $\sigma_v$ of the lens:
\begin{equation}\label{eq:teinsis}
\tein = \frac{\sigma_v^2}{G D_{\mathrm{d}}\Sigma_{\mathrm{cr}}},
\end{equation}
where $D_{\mathrm{d}}$ is the angular diameter distance to the lens.
We set $\zlens=0.3$, $\zsource=1.5$, and assumed a flat $\Lambda$CDM cosmology with $H_0=70\,\rm{km}\,\rm{s}^{-1}\,\rm{Mpc}^{-1}$ and $\Omega_M=0.3$, from which follows
\begin{equation}
\tein \simeq 0.83''\left(\frac{\sigma_v}{200\,\rm{km}\,\rm{s}^{-1}}\right)^2.
\end{equation}
Finally, we assumed that the values of $\sigma_v$ of the galaxy population are drawn from a distribution with the following functional form
\begin{equation}\label{eq:psigmav}
\pr(\sigma_v) \propto \left(\frac{\sigma_v}{\sigma_*}\right)^{\alpha-1}\exp{\left\{-\left(\frac{\sigma_v}{\sigma_*}\right)^\beta\right\}},
\end{equation}
which is often used to describe the velocity dispersion function of galaxies \citep{She++03}.
Based on the measurements of \citet{Mon++17}, we set $\alpha=6.0$, $\beta=2.4$ and $\sigma_*=130\,\rm{km}\,\rm{s}^{-1}$.
Given \Eref{eq:psigmav}, we can compute the probability distribution in quad caustic area of the foreground galaxies, $\prlens(\Aquad)$.
From \Eref{eq:aquadsie} and \Eref{eq:teinsis} it follows that $\Aquad \propto \sigma_v^4$. 
Therefore,
\begin{equation}
\prlens(\Aquad) \propto \Aquad^{-3/4} \pr\left(\sigma_v(\Aquad)\right)
\end{equation}
and
\begin{equation}
\prquad(\Aquad) \propto \Aquad^{1/4} \pr\left(\sigma_v(\Aquad)\right).
\end{equation}
Defining $A_* = \Aquad(\sigma_v = \sigma_*)$, we have
\begin{equation}
\prquad(\Aquad) \propto \Aquad^{(1+\alpha)/4}\exp{\left\{-\left(\frac{\Aquad}{A_*}\right)^{\beta/4}\right\}}
\end{equation}
\Fref{fig:dists} (left-hand panel) shows the distribution in the quad caustic area of the foreground galaxy population and that of the lens population. 
The right-hand panel shows the corresponding distributions in Einstein radius.
These are qualitatively similar to that of existing samples of galaxy-scale lenses \citep[see e.g.][]{Aug++09,Son++13a}.

The last ingredient needed to compute the bias is the likelihood function, $\pr(\Aquadobs|\Aquad)$. We assumed a Gaussian form,
\begin{equation}
\pr(\Aquadobs|\Aquad) = \frac{1}{\sqrt{2\pi}\sigma_A}\exp{\left\{-\frac{(\Aquadobs - \Aquad)^2}{2\sigma_A^2}\right\}},
\end{equation}
centred on the true value and with dispersion $\sigma_A$.
We set $\sigma_A$ to be a fixed fraction $s_A$ of the true value of the quad caustic area,
\begin{equation}\label{eq:sA}
\sigma_A = s_A \Aquad.
\end{equation}
We explored three different values of $s_A$: $0.05, 0.10, 0.20$.

\begin{figure}
\includegraphics[width=\columnwidth]{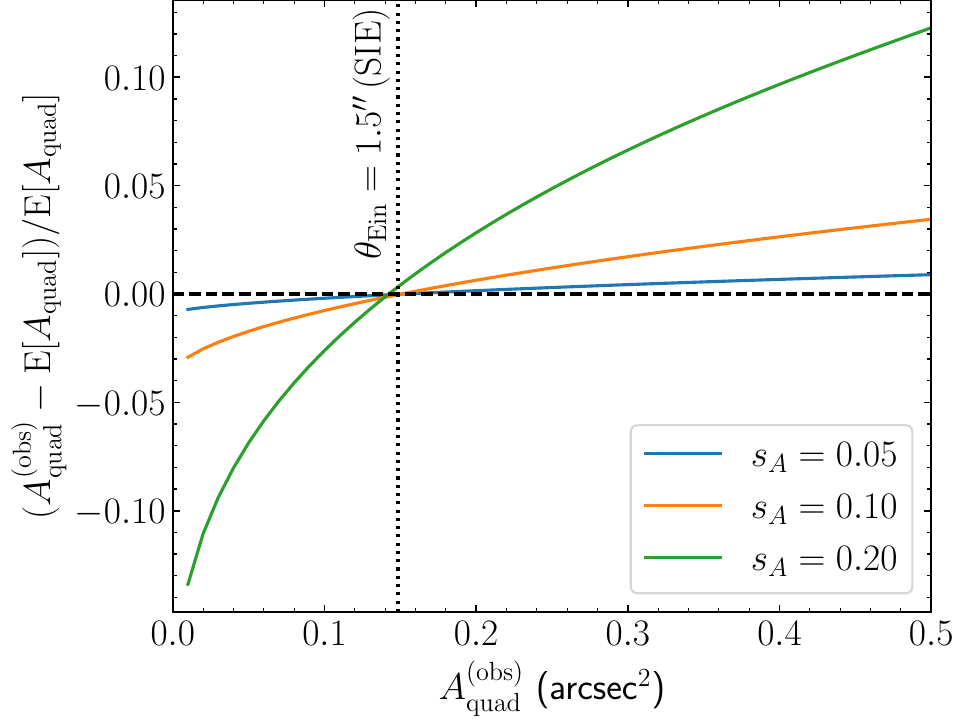}
\caption{
Relative bias on the quad caustic area in our simulation, for three different values of $s_A$, as defined in \Eref{eq:sA}.
The vertical dotted line marks the quad caustic area of an SIE lens with $q=0.7$ and an Einstein radius of $1.5''$, for reference.
\label{fig:bias}
}
\end{figure}

\Fref{fig:bias} shows the relative bias on the quad caustic area, computed using \Eref{eq:bias}, as a function of the observed value.
For small $\Aquadobs$ the bias is negative: the caustic area is underestimated, as predicted by \citet{B+S24}. This is a consequence of the fact that, for a sufficiently large value of the low$-\sigma_v$ slope $\alpha$, $\prquad(\Aquad)$ is an increasing function of $\Aquad$: the observational scatter scrambles the observed values of the caustic area, and there are more lenses with intrinsically larger $\Aquad$ that scatter low than the other way around.
This effect is known as Eddington bias.
However, at large $\Aquadobs$ the trend is the opposite, because $\prquad(\Aquad)$ declines with increasing $\Aquad$.
The transition between negative and positive trend occurs at a larger value of $\Aquadobs$ than the peak in the $\prquad(\Aquad)$ distribution simply because there is more volume in the parameter space at large $\Aquad$, giving more weight to larger values in the integrand of \Eref{eq:bias}.

Our experiment can also give us an idea of the amplitude of the caustic area bias that we can expect in practice.
The key quantity for such an estimate is $s_A$, which describes the fractional scatter in the caustic area measured from lens modelling with respect to the true value.
In the case of time-delay lensing analyses, the dominant source of scatter is the external convergence from line-of-sight matter, $\kappaext$.
To first approximation, the addition of external convergence to a lens model acts like a mass-sheet transformation that re-scales source plane areas by a factor $(1-\kappaext)^2$. Typical values of $\kappaext$ are $\approx0.05$ \citep{Wel++23}, which translate to $s_A\approx0.1$.
According to our simulation, then, the caustic area bias has a typical amplitude of $1-2\%$. 
By the same argument, a $2\%$ bias on $\Aquad$ is equivalent to a $1\%$ bias on $\kappaext$. The effect of $\kappaext$ is that of rescaling time delays by $1-\kappaext$, therefore its bias propagates directly to the inferred value of $H_0$.
A $1\%$ bias on $H_0$ is smaller than the current statistical uncertainty from time-delay analyses.
Moreover,
as we discuss in \Sref{sect:concl}, this is likely an overestimate due to our bright source limit assumption.

We now go back to considering the general case. We argue that, if the likelihood function is known, then the caustic area bias can be accounted for in a lensing study.
The problem of determining the true value of $\Aquad$ of a lens is no different from that of determining any noisy property of a galaxy.
In both cases, the key is to have an accurate prior on the underlying distribution of the parent population of objects and an accurate knowledge of the likelihood function.
For example, assuming a uniform prior on $\Aquad$ leads to a bias of the kind shown in \Fref{fig:bias}, just like assuming a uniform prior on the stellar mass of a galaxy leads to a bias originating from the non-uniformity of the galaxy stellar mass function (i.e. the Eddington bias).
One way of removing the bias is by modelling the underlying population of objects.
In the lensing case, this can be done by fitting a hierarchical model to a large sample of lenses \citep[see e.g.][]{Son++15,Sha++21}.
Most importantly, although the lensing selection determines the distribution in the lens properties $\prquad$, it is perfectly possible to infer the latter without explicitly accounting for the former.
As we show in the next section, however, selection effects can play a role at the lens modelling stage.


\section{Bias from modelling}\label{sect:modelling}

In this section we describe a simple scenario in which a seemingly straightforward lens modelling approach leads to a biased estimate of the lens properties. We show that this is a kind of caustic area bias, originating from an improper treatment of the lens selection procedure, and give a prescription for how to remove it.

\begin{figure*}
\includegraphics[width=\textwidth]{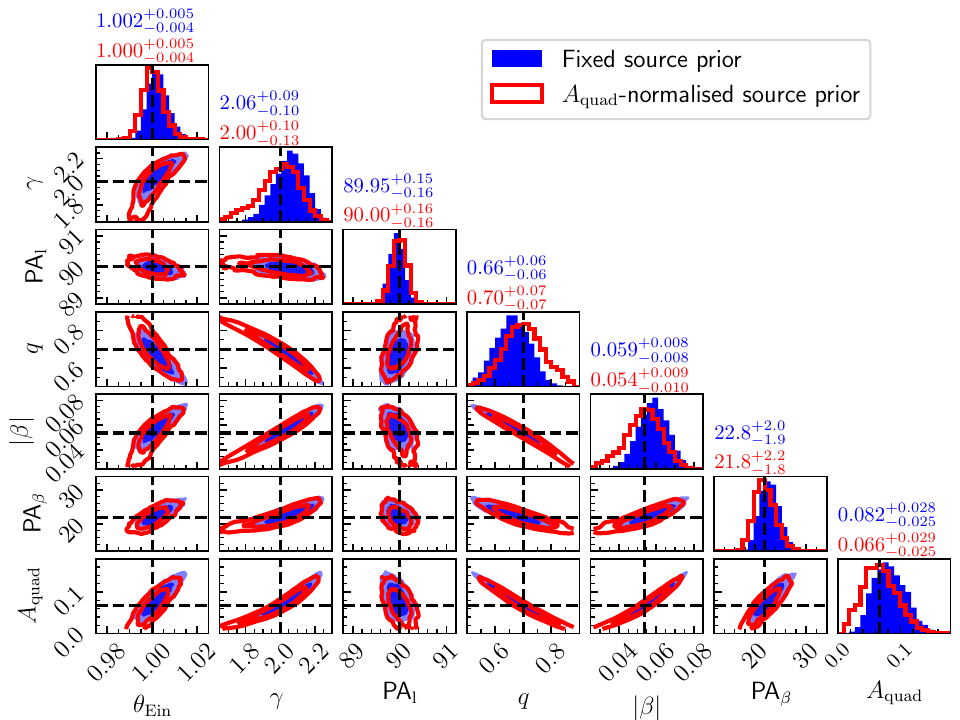}
\caption{
Posterior probability distribution of the power-law ellipsoid fit described in \Sref{sect:modelling}.
Blue filled contours: posterior probability obtained assuming a fixed uniform prior on the source position.
Red solid contours: posterior probability of the model with the prior of \Eref{eq:goodprior}.
The quantities shown are: Einstein radius (in arcsec), density slope, position angle of the major axis (in degrees), axis ratio, source position in polar coordinates (distance from optical axis in arcsec and angle in degrees), and quad caustic area (in square arcsec).
The dashed lines indicate the true values of the parameters.
The median and $68\%$ credible region of the marginal posterior in each quantity is listed at the top of each column.
\label{fig:cp}
}
\end{figure*}

We took an SIE lens with Einstein radius $\tein=1.0''$ and axis ratio $q=0.7$. We placed a background point source located within the tangential caustic, hence producing a quad configuration. 
Then, we fitted a power-law ellipsoid to the four image positions of the lensed source.
We assumed the centroid of the lens to be known exactly and we optimised for the following parameters: the Einstein radius $\tein$, the density slope $\gamma$, the axis ratio $q$, the position angle (PA) of the major axis, and the source position $\boldsymbol\beta$.
We assumed a uniform prior within generous fixed bounds for all of the lens parameters.
We assigned a $0.005''$ uncertainty to the observed image positions, then sampled the posterior probability distribution of the model parameters with a Markov Chain Monte Carlo (MCMC). We used the software {\sc Glafic} \citep{Ogu21} to solve the lens equation for each set of model parameters and {\sc Emcee} \citep{For++13} to run the MCMC.
The number of constraints (eight) exceeds the number of free parameters (six), therefore the problem is well-posed.
The power-law ellipsoid model family used for the fit includes the true values of the lens parameters. Since we did not add noise to the data, one would expect that the fit recovered the input.
However, as \Fref{fig:cp} shows (blue contours), the posterior probability of the individual lens parameters is biased with respect to the truth.
For instance, the median value of the marginalised posterior on the density slope is $\gamma=2.06$, substantially larger than the input value of $\gamma=2.00$.

The source of the bias is the following.
The lens was selected to be a quad, but this piece of information was not used during the fit.
To do it, we must assume the following prior on the source position $\boldsymbol\beta$:
\begin{equation}\label{eq:goodprior}
\pr(\boldsymbol\beta) = \left\{\begin{array}{ll} \dfrac{1}{\Aquad} & \rm{if}\,\boldsymbol\beta\,{is\,within\,the\,quad\,region} \\
0 & \rm{otherwise}\end{array}\right. .
\end{equation}
In words, the prior of \Eref{eq:goodprior} is constant over the region corresponding to a quad configuration (in most practical cases, the region enclosed within the tangential caustic) and zero outside of it: models with fewer than four images are not considered, because such lenses would not have been selected in the first place.
Naively one might think that, since the likelihood of a model with the wrong number of images is zero anyway, there is no practical difference between the original fixed uniform prior and that of \Eref{eq:goodprior}.
However, the two differ under one crucial aspect: the normalisation.
Neglecting it causes models corresponding to larger values of $\Aquad$ to be favoured.
This can be clearly seen in \Fref{fig:cp}: the quad caustic area $\Aquad$ correlates with all of the other lens model parameters, and the inference of these parameters is biased towards values that correspond to a larger $\Aquad$.
\Fref{fig:cp} also shows the inference obtained with the correct prior of \Eref{eq:goodprior}: in this case all of the input values of the parameters are accurately recovered.
This bias looks qualitatively similar to one reported by \citet{R+K24}, who performed a similar experiment to the one described here. It is possible that their bias can also be ascribed to the choice of the source position prior.

Rescaling the source position prior on the basis of the caustic area is not a standard practice in lens modelling. 
This raises the question of why such a large bias on the density slope of power-law models has not been reported before. The answer is in the quality of the data used for the fit. In our experiment we only fitted image positions: these can allow us to measure the Einstein radius of a lens, but have little constraining power on the density profile. Instead, state-of-the-art lens models are obtained by fitting the full surface brightness distribution of an extended source (in the case of a lensed quasar, the host galaxy).
Therefore, we believe that the amplitude of bias is a function of the 
constraining power of the lensing data.
We verified this conjecture by repeating the experiment while fitting flux ratios, in addition to the image positions: we did not detect any bias in that case.
This is in agreement with results from tests of power-law ellipsoid fits to simulated lenses with smooth extended sources \citep{Suy12}.
Tests on more complex simulations, such as those of the time delay lens modelling challenge \citep{Din++21}, have revealed biases on the inferred lens parameters, though they are likely the result of incorrect modelling of the source structure or the point-spread function \citep{Gal++24}.


\section{Discussion and conclusions}\label{sect:concl}

In \Sref{sect:population} we argued that, when the likelihood of the lensing data is known, the caustic area bias can be accounted for by modelling the population of lenses.
In \Sref{sect:modelling} we showed that, although the standard way of setting the source position prior can lead to a biased inference of the parameters of a lens, such a bias can be removed by accounting for the lens selection criteria. We also pointed out how this lens modelling bias disappears in the limit in which the model is well-constrained by the data. 
Although this work deals exclusively with quad lenses, all findings can be generalised to lenses of any configuration.

The entire analysis was carried out in the limit of a very bright source.
Nevertheless, we can easily drop that assumption.
The first step is to introduce the source flux parameter, $\sflux$, and its corresponding population distribution $\prsource(\sflux)$.
Second, we introduce the strong lensing cross-section, as defined by \citet{Son++23}:
\begin{equation}\label{eq:crosssect}
\crosssect = \int_{\mathbb{R}^2} d\boldsymbol\beta \pdet(\psilens,\beta,\sflux|\mathrm{quad}).
\end{equation}
In this equation, $\pdet$ is the probability of detecting a strong lens, given the parameters describing the lens galaxy, $\psilens$, the background source flux and position, and the selection criterion, which in our case is the detection of four images.
Next, in order to repeat the analysis of \Sref{sect:population}, we need to marginalise $\crosssect$ over the source flux, to obtain an effective cross-section averaged over the background source population: 
\begin{equation}
\crosssecteff \equiv \int d\sflux \crosssect\prsource(\sflux).
\end{equation}
At that point we can simply replace $\Aquad$ with $\crosssecteff$.
Similarly, we can generalise the prior of \Eref{eq:goodprior} as follows:
\begin{equation}\label{eq:complexprior}
\pr(\boldsymbol\beta,\sflux) = \frac{1}{\crosssecteff}\pdet(\psilens,\boldsymbol\beta,\sflux)\prsource(\sflux).
\end{equation}

This formalism can allow us in principle to describe scenarios with arbitrarily complex selection criteria, which are encoded in $\pdet$.
The difficult part is obtaining an accurate description of these criteria on real lenses.
Even assuming that $\pdet$ is known exactly, implementing a prior of the kind of \Eref{eq:complexprior} in a lens modelling code might not be easy: it would require, for each possible set of lens model parameters $\psilens$, to compute the lensing cross-section for all possible values of the source flux. 

One important consequence of dropping the bright source assumption is that the amplitude of the caustic area biases decreases.
This was already pointed out by \citet{B+S24}, and follows from the fact that, in all practical cases, increasing the caustic area at fixed Einstein radius causes the magnification to decrease. As a result, the variation of $\crosssecteff$ with lens model parameters is milder than in the bright source case, which means that the likelihood $\pr(\Aquadobs|\Aquad)$ becomes narrower.

In light of our findings, a time-delay cosmography study that aims to remove all sources of caustic area bias must do the following.
First, model individual lenses while ensuring that the lensing data used for this purpose has high signal-to-noise ratio. 
The actual threshold in signal-to-noise ratio needs to be determined via dedicated simulations and depends on the level of accuracy desired.
The end product of the individual modelling phase will be lens models that are constrained to high precision, but that are still potentially subject to the mass-sheet degeneracy, from line-of-sight structure or from a mismatch between the model family and the true density profile of the lenses \citep[such as the power-law models used by][]{Bir++20}.
Second, obtain measurements of an ensemble of observables that correlates with the mass-sheet transformation parameter $\lmst$ at fixed Einstein radius in a well-understood way (again, in relation to the desired level of accuracy). 
Such observables could be the number of galaxy counts around the main lens, which can be related with the external convergence, or the velocity dispersion profiles of the lenses, which constrain their internal structure.
Third, fit a hierarchical model describing how the lens parameters $\psilens$, which now include $\lmst$, are distributed among the population of lenses.
State-of-the-art time delay lensing analyses are already going in this direction: \citet{Bir++20} included a hierarchical model of the lens population in their study of seven time-delay lenses, although only two of the lens model parameters were modelled hierarchically ($\lmst$ and the orbital anisotropy parameter).
Future lens population models should aim at capturing the distribution of all parameters related to the mass density profile of the lenses.
By doing so, the Eddington bias on all quantities that correlate with the time delays, including but not limited to the caustic area, can be taken care of.
In any case, given its very small amplitude, that of removing the caustic area bias is not a priority for current time-delay lensing analyses.

\begin{acknowledgements}
I thank Paul Schechter for his constructive feedback on this manuscript.
This work was supported by the National Science Foundation of China, with an Excellent Young Scholars Fund Overseas grant.
\end{acknowledgements}

\bibliographystyle{aa}
\bibliography{references}

\end{document}